# Quasi-optimal observables: Attaining the quality of maximal likelihood in parameter estimation when only a MC event generator is available

**Fyodor V. Tkachov**

*Institute for Nuclear Research of Russian Academy of Sciences,
7a, 60th October Ave., Moscow, 117312, Russian Federation*

A new method of quasi-optimal observables allows one to approach the quality of data processing usually associated with the method of maximal likelihood within the simpler algorithmic context of generalized moments.

In this lecture, I'd like to explain a recent finding [1] which connects the two basic methods of parameter estimation, the method of maximal likelihood and the method of generalized moments (see e.g. [2]). The two methods (along with the $\chi^2$ method, which I won't discuss) are very well known and widely used in experimental physics.

In a sense, the connection views the method of maximal likelihood as corresponding to a special point in the space of generalized moments, and considers small deviations from that point. The point corresponds to the minimum of the fundamental Cramer-Rao inequality, and small deviations from it introduce non-optimalities (compared with the maximal likelihood method) that are only quadratic in the deviations. This approach offers what appears to be a new and useful algorithmic scheme which combines the theoretical advantage of the method of maximal likelihood (i.e. the fact that it yields the absolute minimum for the variance of the parameter being estimated with a given data sample) with the algorithmic simplicity of the method of moments.

I call the resulting method ***the method of quasi-optimal observables***. It is useful in situations where the method of maximal likelihood fails or cannot be applied, e.g. in high energy physics where typically only a Monte Carlo event generator is available but no explicit formula for the probability density.

One deals with a random variable $\mathbf{P}$ whose instances (specific values) are called events. Their probability density is denoted as $\pi(\mathbf{P})$. It is assumed to depend on a parameter $M$ which has to be estimated from an experimental sample of events $\{\mathbf{P}_i\}_i$.

The method of generalized moments consists in choosing a function $f(\mathbf{P})$ defined on events (the generalized moment or, using the language of quantum theory, observable), and then finding $M$ by fitting its theoretical average value,

$$\langle f \rangle = \int d\mathbf{P}\, \pi(\mathbf{P})\, f(\mathbf{P}), \tag{1}$$

against the corresponding experimental value:

$$\langle f \rangle_{\exp} = \frac{1}{N} \sum_i f(\mathbf{P}_i). \tag{2}$$

The result of the fit is an estimate for $M$ denoted as $M[f]$. Once the observable $f$ is chosen,

the method is rather easy to use. However, the method says nothing about how to find a good $f$, i.e. one which would minimize the variance $D(M[f])$ of the resulting estimate $M[f]$.

The method of maximal likelihood, on the other hand, prescribes to choose $M$ which maximizes the likelihood function

$$L = \prod_i \pi(\mathbf{P}_i). \tag{3}$$

The necessary condition for the minimum then is

$$\frac{\partial L}{\partial M} = L \sum_j \frac{\partial \ln \pi(\mathbf{P}_j)}{\partial M} = 0. \tag{4}$$

The method, if applicable, yields an estimate $M_{\text{opt}}$ for $M$ whose variance $D(M_{\text{opt}})$ is optimal because it is asymptotically equal to the minimal value established by the fundamental Cramer-Rao inequality (cf. Eq. (8.10) in [2]; $N$ is the number of events $\mathbf{P}_i$):

$$D(M_{\text{opt}}) \underset{N \to \infty}{\approx} N^{-1} \left\langle \left(\frac{\partial \ln \pi}{\partial M}\right)^2 \right\rangle^{-1}. \tag{5}$$

Although theoretically ideal, the method of maximal likelihood may be difficult to make use of, e.g. if the number of events is large and/or there is no sufficiently simple regular expression for the probability density $\pi$. The worst case is, of course, when the explicit expression for $\pi$ is unavailable; this case occurs when all one has is a Monte Carlo event generator.

So, on the one hand, there is a simple but non-optimal method of generalized moments. On the other hand, there is a theoretically ideal but cumbersome and often unusable method of maximal likelihood. And there is no apparent connection between the two.

Following [1], let us ask a natural question: is it possible to find an observable $f$ which would minimize $D(M[f])$? If such observable $f_{\text{opt}}$ exists, the corresponding $D(M[f_{\text{opt}}])$ must be directly connected to the r.h.s. of (5).

The trick used in [1] is as follows. Asymptotically,

$$ND(M[f]) \underset{N \to \infty}{\approx} \operatorname{Var} M[f] = \left(\frac{\partial \langle f \rangle}{\partial M}\right)^{-2} \operatorname{Var} f, \tag{6}$$

where $\operatorname{Var} f = \left\langle (f - \langle f \rangle)^2 \right\rangle$. Then it is sufficient to consider $\operatorname{Var} M[f]$ as a numeric function in the functional space of $f$ and to use the apparatus of functional derivatives to study the problem similarly to how one studies minima in ordinary spaces. (A note concerning mathematical rigor: the method is valid under the same conditions as the method of maximal likelihood, and the usual Hilbert norm of mathematical statistics is to be chosen in the space of $f$.) The necessary condition for the minimum is

$$\frac{\delta}{\delta f(\mathbf{P})} \operatorname{Var} M[f] = 0. \tag{7}$$

After simple calculations (see [1] for details) one finds the following solution:

$$f_{\text{opt}}(\mathbf{P}) = \frac{\partial \ln \pi(\mathbf{P})}{\partial M}. \tag{8}$$

(In fact, there is a family of solutions, $f(\mathbf{P}) = C_1 f_{\text{opt}}(\mathbf{P}) + C_2$.) Another simple calculation yields

$$\text{Var}\, M\left[f_{\text{opt}}\right] = \left\langle f_{\text{opt}}^2 \right\rangle^{-1}. \tag{9}$$

In view of (8) and (5) we see that extracting $M$ using the observable (8) is asymptotically equivalent to the method of maximal likelihood.

Once we adopted the viewpoint of analogy with ordinary functions, a natural next step is to consider small deviations from $f_{\text{opt}}$ and their effect on $\text{Var}\, M[f]$. To this end, expand $\text{Var}\, M[f]$ in $f$ around $f_{\text{opt}}$; what we are doing here is a functional analog of the Taylor theorem:

$$\text{Var}\, M[f_{\text{opt}} + \varphi] = \text{Var}\, M[f_{\text{opt}}] + \frac{1}{2} \int \left[\frac{\delta^2 \text{Var}\, M[f]}{\delta f(\mathbf{P})\delta f(\mathbf{Q})}\right]_{f=f_{\text{op}}} \varphi(\mathbf{P})\varphi(\mathbf{Q})\, \mathrm{d}\mathbf{P}\mathrm{d}\mathbf{Q} + \ldots \tag{10}$$

The term which is linear in $\varphi$ does not occur because $f_{\text{opt}}$ satisfies (7). Explicit calculations (see [1] for details) yield:

$$\text{Var}\, M[f_{\text{opt}} + \varphi] = \frac{1}{\left\langle f_{\text{opt}}^2 \right\rangle} + \frac{1}{\left\langle f_{\text{opt}}^2 \right\rangle^3} \left\{ \left\langle f_{\text{opt}}^2 \right\rangle \times \left\langle \bar{\varphi}^2 \right\rangle - \left\langle f_{\text{opt}} \times \bar{\varphi} \right\rangle^2 \right\} + \ldots \tag{11}$$

where $\bar{\varphi} = \varphi - \langle \varphi \rangle$. Non-negativity of the factor in curly braces follows from the standard Schwartz inequality. From the viewpoint of (11), $\varphi$ is small if

$$\frac{\left\langle f_{\text{opt}}^2 \right\rangle \left\langle \bar{\varphi}^2 \right\rangle - \left\langle f_{\text{opt}} \bar{\varphi} \right\rangle^2}{\left\langle f_{\text{opt}}^2 \right\rangle^2} \ll 1. \tag{12}$$

Since the deviation from optimality is quadratic with respect to the deviation of observables from $f_{\text{opt}}$, one realizes that the exact knowledge of the probability distribution $\pi$ is not really necessary: an approximation $f_{\text{quasi}}$ to $f_{\text{opt}}$ in the sense of (12) may be sufficient. Such an approximation could be constructed even using a Monte Carlo generator.

There are several interesting points about this method.

- The usual procedures of imposing cuts on events to enhance the signal/background ratio fully agree with the above prescriptions. Indeed, suppose $\pi = \pi_{\text{bg}} + \pi_{\text{signal}}$, where only the signal contribution depends on $M$. Then

$$f_{\text{opt}} = \frac{\partial_M \pi_{\text{signal}}}{\pi_{\text{bg}} + \pi_{\text{signal}}} \sim \frac{\partial_M \pi_{\text{signal}}}{\pi_{\text{bg}}}. \tag{13}$$

This vanishes where the background is large compared with the signal.

- The optimal observable is localized on events where $\pi$ exhibits the largest variation with respect to the parameter being studies — not where $\pi$ is largest. In addition, such observables may have different signs in different regions of phase space, e.g. in the case of parameters such as masses. Indeed, for $\pi(\mathbf{P}) \propto \dfrac{1}{(M-\mathbf{P})^2 + \Gamma^2}$, the optimal observable with respect to $M$ has the form $f_{M,\mathrm{opt}}(\mathbf{P}) \sim \dfrac{(M-\mathbf{P})}{(M-\mathbf{P})^2 + \Gamma^2}$. Then one has an array of simple shapes to choose from in construction of quasi-optimal observables as shown below:

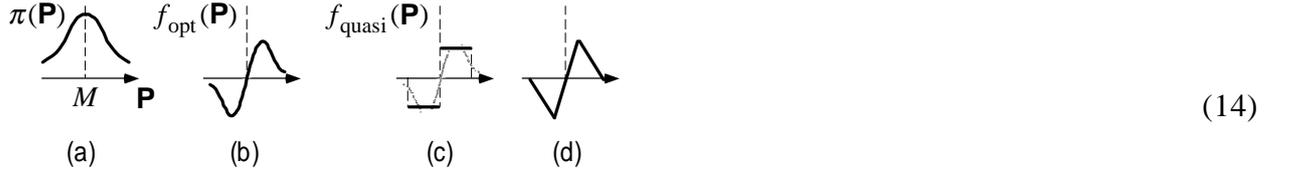

(14)

- In the above example, there is another parameter, $\Gamma$. It is straightforward to define an optimal observable for this parameter too. In general, with several parameters to be estimated, there is an optimal observable per parameter. Error ellipsoids are constructed in the usual fashion.

- A theoretical prediction for $\pi$ may involve a low order result and some higher order corrections. In some cases such corrections will only marginally affect $f_{\mathrm{quasi}}$, so one could construct $f_{\mathrm{quasi}}$ using the simplest expression for $\pi_{\mathrm{theor}}$. However, this affects only the construction of $f_{\mathrm{quasi}}$: once the latter is fixed, the extraction of $M$ from data must involve $\langle f_{\mathrm{quasi}} \rangle$ computed by numerical integration of the $f_{\mathrm{quasi}}$ thus fixed against the theoretical probability distribution with all the corrections taken into account.

- From the algorithmic viewpoint, the problem of numerical construction of a quasi-optimal observable from a MC generator is sister to the problem of MC integration. There is a considerable array of options here (cf. [3]), and given the described firm analytical foundation of the method, I'd expect it to eventually become a tool of choice in many situations where at present less focused methods are used, such as based on neural networks.

To summarize, parameter estimation via quasi-optimal observables combines, within a flexible algorithmic scheme, the optimality of maximal likelihood with the simplicity of generalized moments.

The method of quasi-optimal observables may be useful in experimental situations characterized by:

$\Rightarrow$ high precision requirements and/or low signal;

$\Rightarrow$ many events to be processed and/or the signal not localized sufficiently well for cuts to work;

$\Rightarrow$ a complicated underlying theory (absence of explicit formula for probability distribution $\pi$; complicated higher order corrections; singular theoretical predictions for $\pi$).

Finally, the author is rather uncomfortable with the claim to have discovered a new algorithmic scheme for parameter estimation based on such a simple connection between the two venerable methods — the methods of generalized moments and maximal likelihood — both learnt by O(10000) students worldwide for about half century. However, I checked a large number of textbooks and monographs on mathematical statistics and its applications and failed to find any trace of it being known to the experts. Also, there is an indirect evidence: it is safe to say that *all* known methods of parameter estimation are used in high energy physics one way or another (cf. [4]), and although attempts to construct better observables using e.g. neural networks abound in high energy physics (cf. [4]), there seems to be no trace of the notion of (quasi-) optimal observables being known to high energy physicists. This is utterly puzzling as the connection is so simple. So, if the claim of novelty is correct, the inevitable question is, why the connection was not discovered sooner?

The only explanations I can offer involve history and psychology. Indeed, the geometrical viewpoint of functional analysis was not wide-spread at the time of discovery of the methods of maximal likelihood and generalized moments, and programmers and calculationists still have little working knowledge of it. On top of that, neither students nor researchers feel the perfunctory proofs of elementary textbook results deserve more than a cursory glance: students have so much to learn; mathematicians, so much to prove; data processing experts, so much code to debug. In short, no one can afford to indulge in dwelling upon elementary results when there is so much hard work to be done to earn one's living. In the case of [1], the pattern was broken by an unconventional motivation from the theory of jet observables developed in [5]; the theory ran contrary to some prevailing prejudices, and as is usual in such cases, the author was under pressure to seek all sorts of arguments to fortify it, which led to a foray into the domain of mathematical statistics. Actually, the solution of the old problem of finding optimal jet-finding algorithms described in [5] is per se a sufficient proof (if such were needed) of usefulness of the concept of quasi-optimal observables.

I thank M. Kienzle-Focacci and P. Bhat for their interest. This work was supported in parts by the RFBR grant 99-02-18365 and the NATO grant PST.CLG.977751.

## References

[1] F.V. Tkachov: *Approaching the parameter estimation quality of maximum likelihood via generalized moments*, physics/0001019.

[2] W.T. Eadie, D. Dryard, F.E. James, M. Roos and B. Sadoulet: *Statistical methods in experimental physics*, North-Holland, 1971.

[3] G.P. Lepage: *A new algorithm for adaptive multidimensional integration,* J. Comp. Phys. 27 (1978) 192;
S. Kawabata: *A new Monte-Carlo event generator for high-energy physics,* Comp. Phys. Comm. 41 (1986) 127;
G.I. Manankova, A.F. Tatarchenko and F.V. Tkachov: *MILXy Way: How much better than VEGAS can one integrate in many dimensions?* FERMILAB-Conf-95/213-T;
S. Jadach: *Foam: multi-dimensional general purpose Monte Carlo generator with self-adapting simplectic grid*, physics/9910004.

[4] P.C. Bhat, H. Prosper and S.S. Snyder: *Top quark physics at the Tevatron,* hep-ex/9809011 [Int. J. Mod. Phys. A13 (1998) 5113].

[5] F.V. Tkachov: *Measuring the number of hadronic jets,* Phys. Rev. Lett. 73 (1994) 2405 [hep-ph/9901332];
~: *Measuring multijet structure of hadronic energy flow, or, What is a jet?* hep-ph/9601308 [Int. J. Mod. Phys. A12 (1997) 5411];
~: *A theory of jet definition*, hep-ph/9901444; rev. Jan. 2000.